# Finite-Size Scaling at $\xi/L \gg 1$

Sergio Caracciolo
*Dipartimento di Fisica*
*Università di Lecce and INFN – Sezione di Lecce*
*I-73100 Lecce, ITALIA*
Internet: CARACCIO@LE.INFN.IT

Robert G. Edwards
*Supercomputer Computations Research Institute*
*Florida State University*
*Tallahassee, FL 32306 USA*
Internet: EDWARDS@SCRI.FSU.EDU

Sabino José Ferreira
*Departamento de Física – ICEx*
*Universidade Federal de Minas Gerais*
*Caixa Postal 702*
*Belo Horizonte, MG 30161 BRASIL*
Internet: SABINO@FISICA.UFMG.BR

Andrea Pelissetto[*]
Alan D. Sokal
*Department of Physics*
*New York University*
*4 Washington Place*
*New York, NY 10003 USA*
Internet: PELISSET@MAFALDA.PHYSICS.NYU.EDU, SOKAL@ACF4.NYU.EDU

September 6, 1994

**Abstract**

We present a simple and powerful method for extrapolating finite-volume Monte Carlo data to infinite volume, based on finite-size-scaling theory. We discuss carefully its systematic and statistical errors, and we illustrate it using three examples: the two-dimensional three-state Potts antiferromagnet on the square lattice, and the two-dimensional $O(3)$ and $O(\infty)$ $\sigma$-models. In favorable cases it is possible to obtain reliable extrapolations (errors of a few percent) even when the correlation length is 1000 times larger than the lattice.

---

[*]Address until August 31, 1994. Permanent address: Dipartimento di Fisica and INFN – Sezione di Pisa, Università degli Studi di Pisa, I-56100 Pisa, ITALIA. Internet: PELISSET@SUNTHPI1.DIFI.UNIPI.IT; Bitnet: PELISSET@IPISNSVA.BITNET; Hepnet/Decnet: 39198::PELISSETTO.

No matter how powerful computers get, physicists will always want to study problems that are too difficult for the computers at hand. For example, in statistical mechanics and quantum field theory, physicists want to push to ever larger correlation lengths $\xi$. But Monte Carlo simulations must perforce be carried out on lattices of finite linear size $L$ (limited by computer memory and speed); the data are then extrapolated to the infinite-volume limit $L = \infty$. Obviously this extrapolation — which is based on the theory of finite-size scaling (FSS) [1,2,3] — is feasible in practice only if $\xi/L$ is not too large. But how large?

In this Letter we present a simple and powerful method for performing the extrapolation to $L = \infty$, and discuss carefully its systematic and statistical errors. We illustrate the method using three examples: the two-dimensional three-state Potts antiferromagnet on the square lattice [4], and the two-dimensional $O(3)$ and $O(\infty)$ $\sigma$-models [5,6]. We have found — much to our surprise — that in favorable cases it is possible to obtain reliable extrapolations (errors of a few percent) at $\xi/L$ as large as 10–1000.

Consider, for starters, a model controlled by a renormalization-group (RG) fixed point having *one* relevant operator. Let us work on a periodic lattice of linear size $L$. Let $\xi(\beta, L)$ be a suitably defined finite-volume correlation length [7], and let $\mathcal{O}$ be any long-distance observable (e.g. the correlation length or the susceptibility). Then finite-size-scaling theory [1,2,3] predicts that [9]

$$\frac{\mathcal{O}(\beta, L)}{\mathcal{O}(\beta, \infty)} = f_{\mathcal{O}}\bigl(\xi(\beta, \infty)/L\bigr) + O\bigl(\xi^{-\omega}, L^{-\omega}\bigr), \qquad (1)$$

where $f_{\mathcal{O}}$ is a universal function and $\omega$ is a correction-to-scaling exponent. It follows that if $s$ is any fixed scale factor (usually we take $s = 2$), then

$$\frac{\mathcal{O}(\beta, sL)}{\mathcal{O}(\beta, L)} = F_{\mathcal{O}}\bigl(\xi(\beta, L)/L\bigr) + O\bigl(\xi^{-\omega}, L^{-\omega}\bigr), \qquad (2)$$

where $F_{\mathcal{O}}$ can be expressed in terms of $f_{\mathcal{O}}$ and $f_\xi$.

Our method proceeds as follows [10]: Make Monte Carlo runs at numerous pairs $(\beta, L)$ and $(\beta, sL)$. Plot $\mathcal{O}(\beta, sL)/\mathcal{O}(\beta, L)$ versus $\xi(\beta, L)/L$, using those points satisfying both $\xi(\beta, L) \geq$ some value $\xi_{min}$ and $L \geq$ some value $L_{min}$. If all these points fall with good accuracy on a single curve — thus verifying the Ansatz (2) for $\xi \geq \xi_{min}$, $L \geq L_{min}$ — choose a smooth fitting function $F_{\mathcal{O}}$. Then, using the functions $F_\xi$ and $F_{\mathcal{O}}$, extrapolate the pair $(\xi, \mathcal{O})$ successively from $L \to sL \to s^2 L \to \ldots \to \infty$.

We have chosen to use functions $F_{\mathcal{O}}$ of the form

$$F_{\mathcal{O}}(x) = 1 + a_1 e^{-1/x} + a_2 e^{-2/x} + \ldots + a_n e^{-n/x}. \qquad (3)$$

This form is partially motivated by theory, which tells us that $F(x) \to 1$ exponentially fast as $x \to 0$ [14]. Typically a fit of order $3 \leq n \leq 9$ is sufficient; we increase $n$ until the $\chi^2$ of the fit becomes essentially constant. The resulting $\chi^2$ value provides a check on the systematic errors arising from corrections to scaling and/or from the inadequacies of the form (3).

The *statistical* error on the extrapolated value of $\mathcal{O}_\infty(\beta) \equiv \mathcal{O}(\beta, \infty)$ comes from three sources:

(i) Error on $\mathcal{O}(\beta, L)$, which gets multiplicatively propagated to $\mathcal{O}_\infty$.

(ii) Error on $\xi(\beta, L)$, which affects the argument $x \equiv \xi(\beta, L)/L$ of the scaling functions $F_\xi$ and $F_\mathcal{O}$.

(iii) Statistical error in our estimate of the coefficients $a_1, \ldots, a_n$ in $F_\xi$ and $F_\mathcal{O}$.

The errors of type (i) and (ii) depend on the statistics available at the single point $(\beta, L)$, while the error of type (iii) depends on the statistics in the whole set of runs. Errors (i)+(ii) [resp. (i)+(ii)+(iii)] can be quantified by performing a Monte Carlo experiment in which the input data at $(\beta, L)$ [resp. the whole set of input data] are varied randomly within their error bars and then extrapolated [15,16,17].

The discrepancies between the extrapolated values from different lattice sizes at the same $\beta$ — to the extent that these exceed the estimated statistical errors — indicate the presence of systematic errors and thus the necessity of increasing $L_{min}$ and/or $\xi_{min}$ and/or $n$.

A figure of (de)merit of the method is the relative variance on the extrapolated value $\mathcal{O}_\infty(\beta)$, multiplied by the computer time needed to obtain it [18]. We expect this *relative variance-time product* [for errors (i)+(ii) only] to scale as

$$\text{RVTP}(\beta, L) \approx \xi_\infty(\beta)^{d+z_{int,\mathcal{O}}} G_\mathcal{O}\bigl(\xi_\infty(\beta)/L\bigr) , \qquad (4)$$

where $d$ is the spatial dimension and $z_{int,\mathcal{O}}$ is the dynamic critical exponent of the Monte Carlo algorithm being used; here $G_\mathcal{O}$ is a combination of several static and dynamic finite-size-scaling functions, and depends both on the observable $\mathcal{O}$ and on the algorithm but not on the scale factor $s$. As $\xi_\infty/L$ tends to zero, we expect $G_\mathcal{O}$ to diverge as $(\xi_\infty/L)^{-d}$ (it is wasteful to use a lattice $L \gg \xi_\infty$). As $\xi_\infty/L$ tends to infinity, we expect $G_\mathcal{O} \sim (\xi_\infty/L)^p$ [19], but *the power $p$ can be either positive or negative*. If $p > 0$, there is an optimum value of $\xi_\infty/L$; this determines the best lattice size at which to perform runs for a given $\beta$. If $p < 0$, it is most efficient to use the *smallest* lattice size for which the corrections to scaling are negligible compared to the statistical errors. [This neglects errors of type (iii); the optimization becomes much more complicated if they are included.]

Our first example [4] is the two-dimensional three-state Potts antiferromagnet on the square lattice, which is believed to have a critical point at $\beta = \infty$ [20]. We used the Wang-Swendsen-Kotecký cluster algorithm [21], which appears to have *no* critical slowing-down ($\tau_{int,\mathcal{M}^2_{stagg}} < 5$ uniformly in $\beta$ and $L$) [4]. We ran on lattices $L = 32, 64, 128, 256, 512, 1024, 1536$ at 153 different pairs $(\beta, L)$ in the range $5 \lesssim \xi_\infty \lesssim 20000$. Each run was between $2 \times 10^5$ and $2.2 \times 10^7$ iterations, and the total CPU time was modest by our standards (about 2 years on an IBM RS-6000/370). We took $\xi_{min} = 10$ and $L_{min} = 64$ and used a quintic fit in (3); the result for $F_\xi$ is shown in Figure 1 ($\chi^2 = 56.47$, 84 DF, level = 99%). The extrapolated values from different lattice sizes at the same $\beta$ agree within the estimated statistical errors ($\chi^2 = 35.79$, 93 DF, level > 99%): see Table 1 for an example. The result for $G_\xi$ is shown in Figure 2; the errors are roughly constant for $\xi_\infty/L \gtrsim 0.4$ but rise sharply for smaller $\xi_\infty/L$. The theoretical exponent $p$ computed [19] from the fitted $F_\xi$ is equal to $p = -0.09 \pm 0.06$; the curve suggests that $p = 0$. In practice we were able to obtain $\xi_\infty$ to an accuracy of about 1% (resp. 2%, 3%, 5%) at $\xi_\infty \approx 1000$ (resp. 2000, 5000, 10000).

Next let us consider [5,6] the two-dimensional $O(3)$ $\sigma$-model. We used the Wolff embedding algorithm with standard Swendsen-Wang updates [22,23,8]; again critical slowing-down appears to be completely eliminated. We ran on lattices $L = 32, 48, 64, 96, 128, 192, 256, 384, 512$ at 169 different pairs $(\beta, L)$ in the range $20 \lesssim \xi_\infty \lesssim 10^5$. Each run was between $10^5$ and $5 \times 10^6$ iterations, and the total CPU time was 7 years on an IBM RS-6000/370. We took $\xi_{min} = 20$ and used an eighth-order fit. There appear to be weak corrections to scaling (of order $\lesssim 1.5\%$) in the region $0.3 \lesssim \xi_L/L \lesssim 0.7$ for lattices with $L \lesssim 64$–96. We therefore chose conservatively $L_{min} = 128$ for $\xi_L/L \leq 0.7$, and $L_{min} = 96$ for $\xi_L/L > 0.7$. The result for $F_\xi$ is shown in Figure 3 ($\chi^2 = 38.62$, 62 DF, level = 99%). The result for $G_\xi$ is shown in Figure 4; at large $\xi_\infty/L$ it *decreases* sharply, with a power $p \approx -2$ in agreement with theory [19]. In practice we obtained $\xi_\infty$ to an accuracy of about 0.8% (resp. 1.4%, 2.1%) at $\xi_\infty \approx 10^3$ (resp. $10^4$, $10^5$).

We also carried out a "simulated Monte Carlo" experiment for the $O(N)$ $\sigma$-model at $N = \infty$, by generating data from the exact finite-volume solution plus random noise of 0.1% for $L = 64, 96, 128$, 0.2% for $L = 192, 256$ and 0.5% for $L = 384, 512$ [which is the order of magnitude we attain in practice for $O(3)$]. We considered 35 values of $\beta$ in the range $20 \lesssim \xi_\infty \lesssim 10^6$. We used $\xi_{min} = 20$ and $L_{min} = 64$ (in fact much smaller values could have been used, as corrections to scaling are here very small) and a ninth-order fit; for two different data sets we get $\chi^2 = 114$ (resp. 118) with 166 DF. In practice we obtain $\xi_\infty$ with an accuracy of 0.6% (resp. 1.2%, 2%, 3%) at $\xi_\infty \approx 10^3$ (resp. $10^4$, $10^5$, $10^6$). Here we can also compare the extrapolated values $\xi_\infty^{extr}(\beta)$ with the exact values $\xi_\infty^{exact}(\beta)$. Defining $\mathcal{R} = \sum_\beta [\xi_\infty^{extr}(\beta) - \xi_\infty^{exact}(\beta)]^2/\sigma^2(\beta)$, we find for the two data sets $\mathcal{R} = 17.19$ (resp. 25.81) with 35 DF. Only 6 (resp. 9) points differ from the exact value more than one standard deviation, and none by more than two.

Details on all of these models will be reported separately [4,6].

The method is easily generalized to a model controlled by an RG fixed point having $k$ relevant operators. It suffices to choose $k - 1$ dimensionless ratios of long-distance observables, call them $R = (R_1, \ldots, R_{k-1})$; then the function $F_\mathcal{O}$ will depend parametrically on $R(\beta, L)$. In practice one can divide $R$-space into "slices" within which $F_\mathcal{O}$ is empirically constant within error bars, and perform the fit (3) within each slice. We have used this approach to study the mixed isovector/isotensor $\sigma$-model, taking $R$ to be the ratio of isovector to isotensor correlation length [5,6].

The method can also be applied to extrapolate the exponential correlation length (inverse mass gap). For this purpose one must work in a system of size $L^{d-1} \times T$ with $T \gg \xi_{exp}(\beta, L)$ (compare [11]).

We wish to thank Martin Hasenbusch and especially Jae-Kwon Kim for sharing their data with us, and for challenging us to push to ever larger values of $\xi/L$. The authors' research was supported by CNR, INFN, CNPq, FAPEMIG, DOE contracts DE-FG05-85ER250000 and DE-FG05-92ER40742, NSF grant DMS-9200719, and NATO CRG 910251.

# References

[1] M.N. Barber, in *Phase Transitions and Critical Phenomena*, vol. 8, ed. C. Domb and J.L. Lebowitz (Academic Press, London, 1983).

[2] J.L. Cardy, ed., *Finite-Size Scaling* (North-Holland, Amsterdam, 1988).

[3] V. Privman, ed., *Finite Size Scaling and Numerical Simulation of Statistical Systems* (World Scientific, Singapore, 1990).

[4] S.J. Ferreira and A.D. Sokal, Antiferromagnetic Potts models on the square lattice,
NYU preprint NYU-TH-94/05/01 (May 1994), `hep-lat@ftp.scri.fsu.edu` #9405015; and in preparation.

[5] S. Caracciolo, R.G. Edwards, A. Pelissetto and A.D. Sokal, Phys. Rev. Lett. **71**, 3906 (1993).

[6] S. Caracciolo, R.G. Edwards, A. Pelissetto and A.D. Sokal, in preparation.

[7] It is necessary to choose a definition of $\xi$ which makes sense in a fully finite system. We use the second-moment correlation length defined by equations (4.11)–(4.13) of Ref. [8].

[8] S. Caracciolo, R.G. Edwards, A. Pelissetto and A.D. Sokal, Nucl. Phys. **B403**, 475 (1993).

[9] This form of finite-size scaling assumes hyperscaling, and thus is expected to hold only below the upper critical dimension of the model. See e.g. [3, Chapter I, section 2.7].

[10] Our method has many features in common with those of Lüscher, Weisz and Wolff [11] and Kim [12]. In particular, all these methods share the property of working only with observable quantities ($\xi$, $\mathcal{O}$ and $L$) and not with bare quantities ($\beta$). Therefore, they rely only on "scaling" and not on "asymptotic scaling"; and they differ from other FSS-based methods such as phenomenological renormalization [13].

[11] M. Lüscher, P. Weisz and U. Wolff, Nucl. Phys. **B359**, 221 (1991).

[12] J.-K. Kim, Phys. Rev. Lett. **70**, 1735 (1993); Nucl. Phys. B (Proc. Suppl.) **34**, 702 (1994); University of Arizona preprints AZPH-TH/93-49 and AZPH-TH/94-15.

[13] P. Nightingale, Proc. Konink. Ned. Akad. **B82**, 235, 245, 269 (1978).

[14] H. Neuberger, Phys. Lett. **B233**, 183 (1989); S. Caracciolo and A. Pelissetto, in preparation.

[15] In principle, $\xi$ and $\mathcal{O}$ should be generated from a *joint* Gaussian with the correct covariance. We ignored this subtlety and simply generated *independent* fluctuations on $\xi$ and $\mathcal{O}$.

[16] Errors of type (i) and (ii) can be exactly computed in terms of $f_\xi$ and $f_\mathcal{O}$. For $\xi$ we have $\Delta\xi_\infty/\xi_\infty = K_\xi(\xi_\infty/L)\,\Delta\xi_L/\xi_L$, where $\Delta\xi_L$ (resp. $\Delta\xi_\infty$) is the standard deviation on the raw (resp. extrapolated) value, and $K_\xi(z) = f_\xi(z)/[f_\xi(z) + z f'_\xi(z)]$. For a derivation, see [6].

[17] The extrapolated estimates from different $(\beta, L)$ are correlated. In setting the error bars we have kept account of the full covariance matrix between extrapolations at the same $\beta$ but different $L$; but we have ignored correlations between extrapolations at different values of $\beta$.

[18] This variance-time product tends to a constant as the CPU time tends to infinity. However, if the CPU time used is too small, then the variance-time product can be significantly larger than its asymptotic value, due to nonlinear cross terms between error sources (i) and (ii).

[19] The exponent $p$, which does not depend on $s$, can be computed in terms of $F_\xi$. Define an exponent $h$ by assuming that $\Delta\xi_L/\xi_L \approx \xi_\infty^z H(\xi_\infty/L)/N_{iter} \sim \xi_\infty^{z+h}/N_{iter}$, where $\Delta\xi_L$ is the error on $\xi_L$ for a run of length $N_{iter}$. If $\xi_L/L \to x_* < \infty$ when $\xi_\infty/L \to \infty$ (this is the case for models with critical exponent $\eta > 0$) and $R \equiv 1 + x_* F'_\xi(x_*)/s > 1$, then $p = -d + 2\log R/\log s + 2h$. If $R = 1$, logarithmic terms appear and $G_\xi(z) \sim z^{-d+2h}(\log z)^q$ with $q > 0$. If instead $\xi_L/L \to \infty$ when $\xi_\infty/L \to \infty$ (this is the case for asymptotically free theories), then $G_\xi(z) \sim z^{-d+2h}(\log z)^2$.

[20] R.J. Baxter, Proc. Roy. Soc. London **A383**, 43 (1982).

[21] J.-S. Wang, R.H. Swendsen and R. Kotecký, Phys. Rev. Lett. **63**, 109 (1989) and Phys. Rev. **B42**, 2465 (1990); M. Lubin and A.D. Sokal, Phys. Rev. Lett. **71**, 1778 (1993).

[22] U. Wolff, Phys. Rev. Lett. **62**, 361 (1989).

[23] R.H. Swendsen and J.-S. Wang, Phys. Rev. Lett. **58**, 86 (1987).

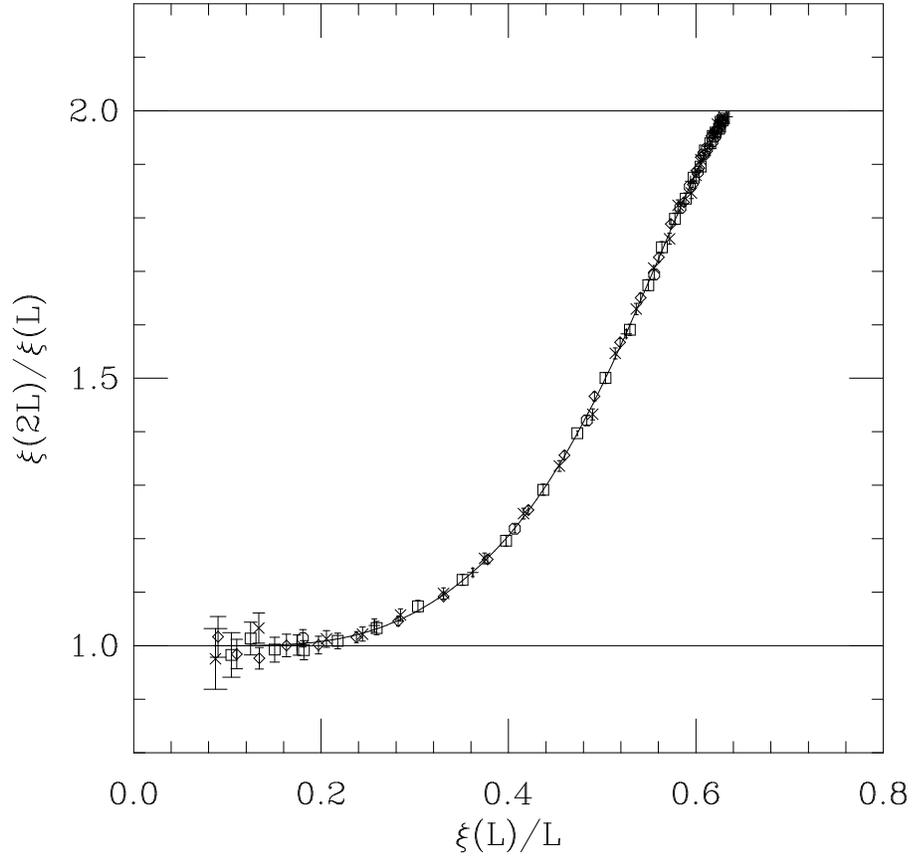

Figure 1: $\xi(\beta, 2L)/\xi(\beta, L)$ versus $\xi(\beta, L)/L$ for the two-dimensional three-state Potts antiferromagnet. Symbols indicate $L = 32$ (+), 64 (×), 128 (□), 256 (◇), 512 (○). Error bars are one standard deviation. Curve is a quintic fit in (3), with $\xi_{min} = 10$ and $L_{min} = 64$.

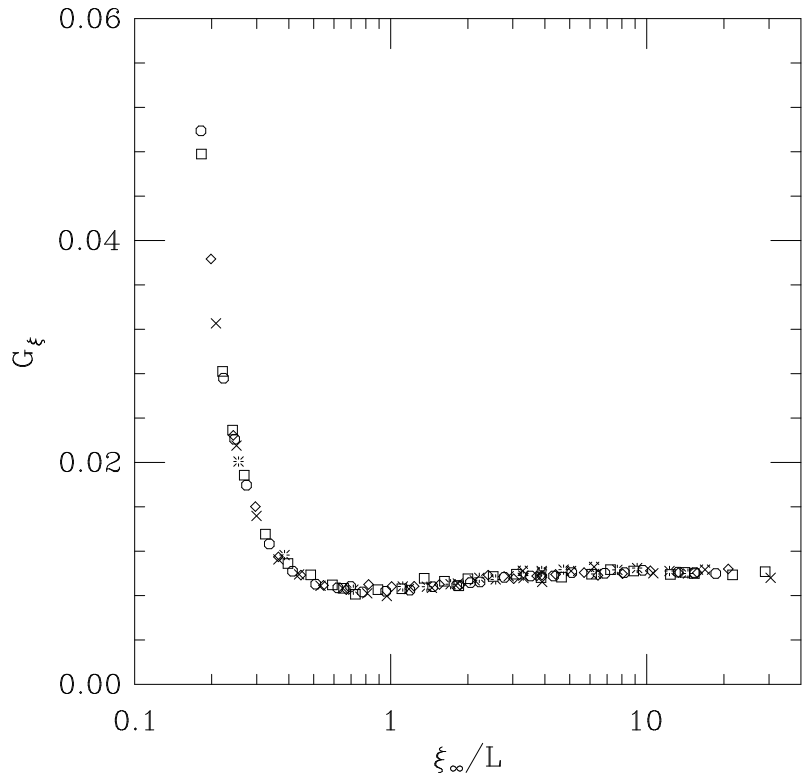

Figure 2: Relative variance-time product [for errors (i)+(ii) only] divided by $\xi_\infty(\beta)^2$, plotted versus $\xi_\infty(\beta)/L$, for two-dimensional three-state Potts antiferromagnet. Symbols indicate $L = 32$ (+), 64 ($\times$), 128 ($\square$), 256 ($\diamond$), 512 ($\circ$), 1024 ($*$), 1536($\boxplus$).

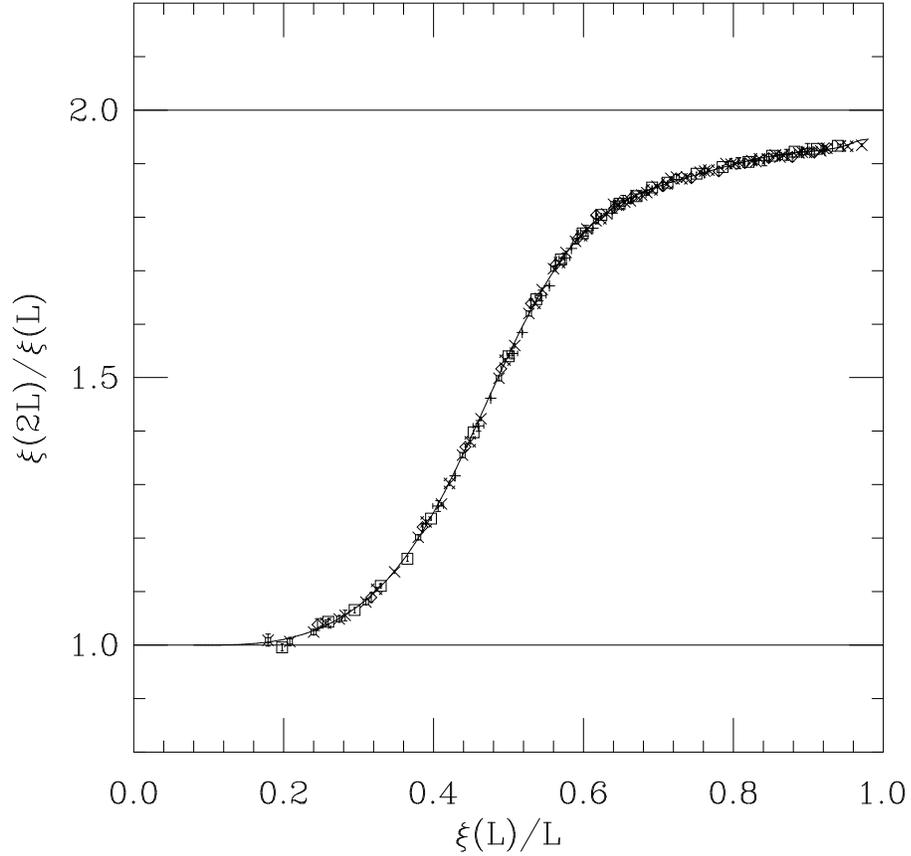

Figure 3: $\xi(\beta,2L)/\xi(\beta,L)$ versus $\xi(\beta,L)/L$ for the two-dimensional $O(3)$ $\sigma$-model. Symbols indicate $L = 32$ (+), 48 (⊞), 64 (×), 96 (※), 128 (□), 192 (⋈), 256 (◇). Error bars are one standard deviation. Curve is an eighth-order fit in (3), with $\xi_{min} = 20$ and $L_{min} = 128$ (resp. 96) for $\xi(L)/L \leq 0.7$ (resp. $> 0.7$).

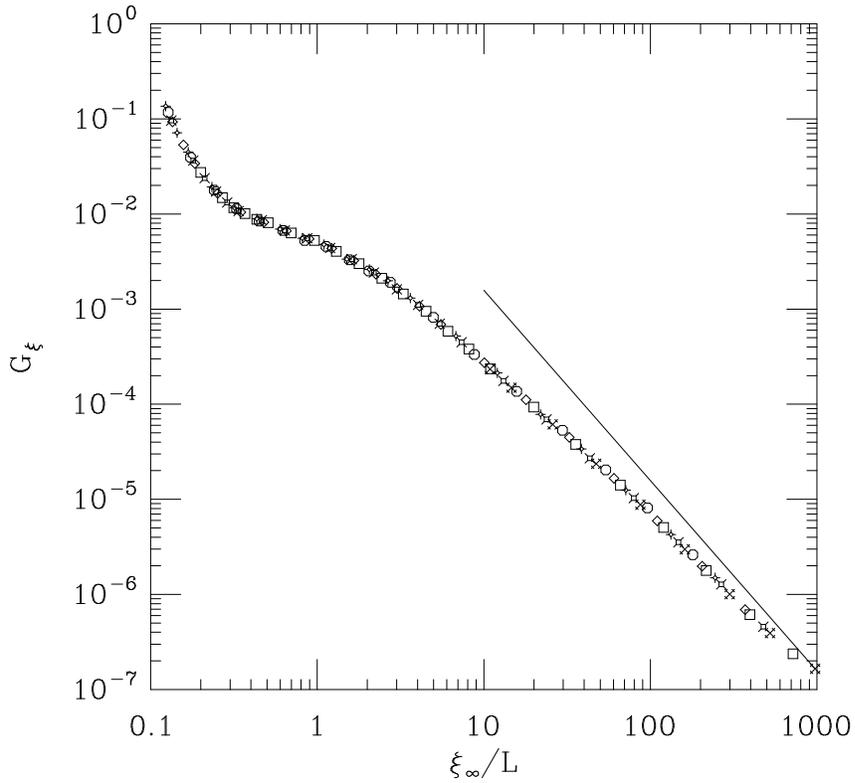

Figure 4: Relative variance-time product [for errors (i)+(ii) only] divided by $\xi_\infty(\beta)^2$, plotted versus $\xi_\infty(\beta)/L$, for two-dimensional $O(3)$ $\sigma$-model. Symbols indicate $L =$ 32 (+), 48 (⊞), 64 (×), 96 (✵), 128 (□), 192 (⌶), 256 (◇), 384 (⟡), 512 (○). For comparison, the line shows the theoretical limiting slope $-2$.

| $L$ | Iterations | Raw Data $\xi(\beta, L)$ | Extrapolated $\xi(\beta, \infty)$ |
|---:|---|---|---|
| 32 | $2.2 \times 10^7$ | 19.02 (0.01) | 90.51 ( 1.01) |
| 64 | $10^6$ | 35.52 (0.04) | 92.66 ( 0.79) |
| 128 | $10^6$ | 60.61 (0.09) | 93.17 ( 0.42) |
| 256 | $10^6$ | 84.69 (0.15) | 93.19 ( 0.30) |
| 512 | $5 \times 10^5$ | 92.48 (0.33) | 92.89 ( 0.38) |
| 1024 | $2 \times 10^5$ | 93.78 (1.17) | 93.78 ( 1.16) |
| mean ($L \geq 64$) | | $\chi^2 = 0.85$ (4 DF, level = 93%) | 93.13 ( 0.26) |

Table 1: Raw and extrapolated correlation lengths for the two-dimensional three-state Potts antiferromagnet at $\beta = 3.5$. Extrapolation based on $\xi_{min} = 10$ and $L_{min} = 64$ and a quintic fit. For each extrapolated value we have reported the standard deviation of the estimate, including errors of all three types. The mean value and the $\chi^2$ have been computed taking into account the full covariance matrix [17].